\documentclass[11pt]{cernrep}
\usepackage{epsfig}
\usepackage{graphicx}
\usepackage{cite,./mcite}

\newcommand{\asmz}{\alpha_s(M_Z^2)}
\newcommand{\msbar}{\mbox{$\overline{\rm{MS}}$}\ }

\begin{document}
\title{Comparison and combination of ZEUS and H1 PDF analyses \\
HERA - LHC Workshop Proceedings }
\author{Amanda Cooper-Sarkar, Claire Gwenlan}
\institute{Oxford University}
\maketitle
\begin{abstract}
The H1 and ZEUS published PDF analyses are compared, including a discussion of 
the different treatments of correlated systematic uncertainties. Differences 
in the data sets and the analyses are investigated by putting the H1 data set 
through both PDF analyses and by putting the ZEUS and H1 data sets through the 
same (ZEUS) analysis, separately. Finally, the HERA averaged data set is put 
through the ZEUS PDF analysis and the result is compared to that obtained when 
putting the ZEUS and H1 data sets through this analsysis together, 
using both the Offset and Hessian methods of treating correlated 
systematic uncertainties.
\end{abstract}


Parton Density Function (PDF) determinations are usually global 
fits~\cite{mrst,cteq,zeus-s}, which use fixed target 
DIS data as well as HERA data. In such analyses the high statistics HERA NC 
$e^+p$ data, which span the range $6.3 \times 10^{-5} < x < 0.65,
2.7 < Q^2 < 30,000$GeV$^2$, 
have determined the low-$x$ sea and 
gluon distributions, whereas the fixed target data have determined 
the valence distributions and the higher-$x$ sea distributions. 
The $\nu$-Fe fixed target data have been the most important input  
for determining the valence distributions, but these data suffer 
from uncertainties due to heavy target corrections. Such uncertainties 
are also present for deuterium fixed target data, 
which have been used to determine the shape of the high-$x$ $d$-valence quark.

HERA data on neutral and charged current (NC and CC) 
$e^+p$ and $e^-p$ inclusive double differential 
cross-sections are now available, 
and have been used by both the H1 and ZEUS collaborations~\cite{zeusj,*h1} 
in order to determine the parton distributions functions (PDFs) using data 
from within a single experiment. The HERA high $Q^2$ cross-section 
data can be used to determine the valence 
distributions, thus eliminating uncertainties from heavy target corrections. 
The PDFs are presented with full accounting for uncertainties from correlated 
systematic errors (as well as from statistical and uncorrelated sources).
Peforming an analysis within a single experiment has considerable advantages
in this respect, since the global fits 
have found significant tensions between 
different data sets, which make a rigorous statistical treatment of 
uncertainties difficult. 

\begin{figure}[tbp]
\centerline{
\epsfig{figure=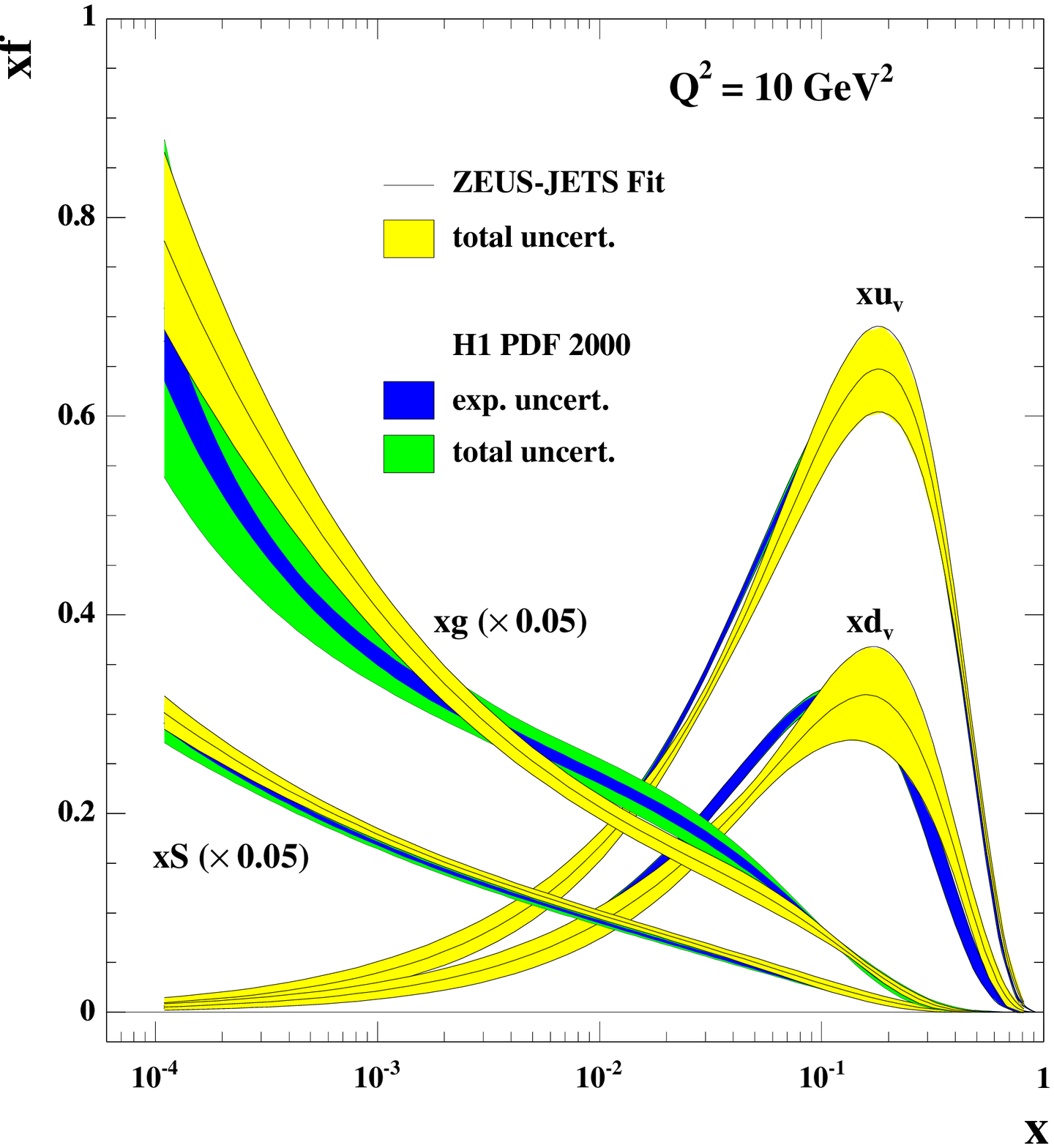,height=6cm}
\epsfig{figure=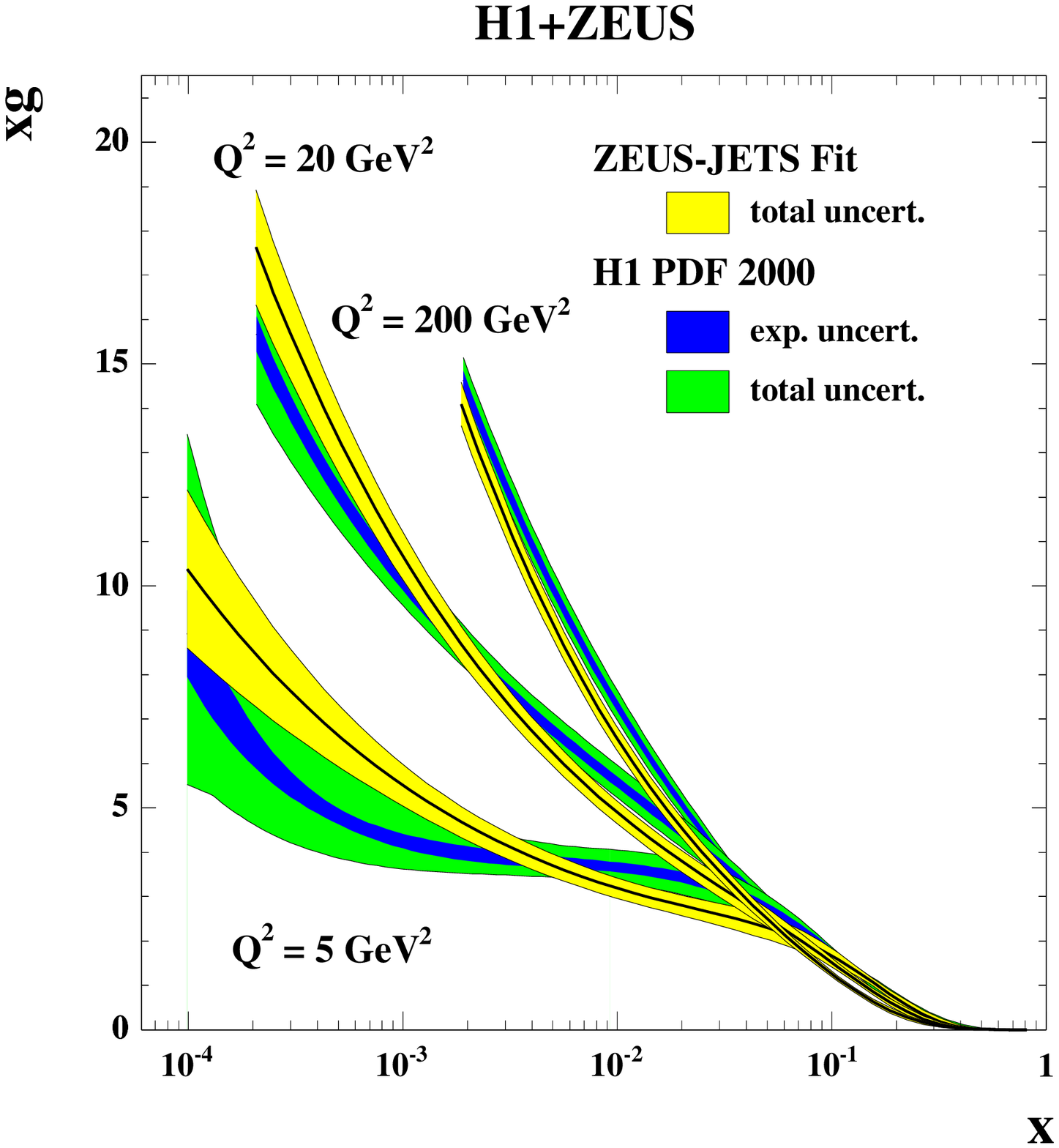,height=6cm}}
\caption {Left plot: Comparison of PDFs from ZEUS and H1 analyses at $Q^2=10$GeV$^2$.
Right plot: Comparison of gluon from ZEUS and H1 analyses, at various $Q^2$. 
Note that the ZEUS analysis total uncertainty includes both experimental and 
model uncertainties.}
\label{fig:h1zeus}
\end{figure}

Fig.~\ref{fig:h1zeus} compares the results of the H1 and ZEUS analyses. 
Whereas the extracted PDFs are broadly compatible within errors, there is a 
noticeable difference in the shape of the gluon PDFs.
Full details of the analyses are given in the relevant publications,
in this contribution we examine the differences in the two analyses, recapping 
only salient details.

The kinematics
of lepton hadron scattering is described in terms of the variables $Q^2$, the
invariant mass of the exchanged vector boson, Bjorken $x$, the fraction
of the momentum of the incoming nucleon taken by the struck quark (in the 
quark-parton model), and $y$ which measures the energy transfer between the
lepton and hadron systems.
The differential cross-section for the NC process is given in terms of the
structure functions by
\[
\frac {d^2\sigma(e^{\pm}p) } {dxdQ^2} =  \frac {2\pi\alpha^2} {Q^4 x}
\left[Y_+\,F_2(x,Q^2) - y^2 \,F_L(x,Q^2)
\mp Y_-\, xF_3(x,Q^2) \right],
\]
where $\displaystyle Y_\pm=1\pm(1-y)^2$. 
The structure functions $F_2$ and $xF_3$ are 
directly related to quark distributions, and their
$Q^2$ dependence, or scaling violation, 
is predicted by pQCD. At $Q^2 \leq 1000$~GeV$^2$ $F_2$ dominates the
charged lepton-hadron cross-section and for $x \leq 10^{-2}$, $F_2$ itself 
is sea quark dominated but its $Q^2$ evolution is controlled by
the gluon contribution, such that HERA data provide 
crucial information on low-$x$ sea-quark and gluon distributions.
At high $Q^2$, the structure function $xF_3$ becomes increasingly important, 
and
gives information on valence quark distributions. The CC interactions 
enable us to separate the flavour of the valence distributions 
at high-$x$, since their (LO) cross-sections are given by, 
\[
\frac {d^2\sigma(e^+ p) } {dxdQ^2} = \frac {G_F^2 M_W^4} {(Q^2 +M_W^2)^2 2\pi x}
x\left[(\bar{u}+\bar{c}) + (1 - y)^2 (d + s) \right],
\]
\[
\frac {d^2\sigma(e^- p) } {dxdQ^2} = \frac {G_F^2 M_W^4} {(Q^2 +M_W^2)^2 2\pi x}
x\left[(u + c) + (1 - y)^2 (\bar{d} + \bar{s}) \right].
\]
For both HERA analyses the QCD predictions for the structure functions 
are obtained by solving the DGLAP evolution equations~\cite{ap,*gl,*l,*d} 
at NLO in the \msbar scheme with the
renormalisation and factorization scales chosen to be $Q^2$. 
These equations yield the PDFs
 at all values of $Q^2$ provided they
are input as functions of $x$ at some input scale $Q^2_0$. 
The resulting PDFs are then convoluted with coefficient functions, to give the
structure functions which enter into the expressions for the cross-sections.
For a full explanation of the relationships between DIS cross-sections, 
structure functions, PDFs and the QCD improved parton model see 
ref.~\cite{dcs}.

The HERA data are all in a kinematic region where there is no
sensitivity to target mass and higher 
twist contributions but a minimum $Q^2$ cut must be imposed 
to remain in the kinematic region where
perturbative QCD should be applicable. For ZEUS this is $Q^2 > 2.5$~GeV$^2$, 
and for H1 it is $Q^2 > 3.5$~GeV$^2$. Both collaborations have included the 
sensitivity to this cut as part of their model errors.

In the ZEUS analysis, the PDFs for $u$ valence, $xu_v(x)$,  $d$ valence, $xd_v(x)$, 
total sea, $xS(x)$, the 
gluon, $xg(x)$, and the difference between the $d$ and $u$
contributions to the sea, $x(\bar{d}-\bar{u})$, are each parametrized  
by the form 
\begin{equation}
  p_1 x^{p_2} (1-x)^{p_3} P(x),
\label{eqn:pdf}
\end{equation}
where $P(x) = 1 +p_4 x$, at $Q^2_0 = 7$GeV$^2$. The total sea 
$xS=2x(\bar{u} +\bar{d} +\bar{s}+ \bar{c} +\bar{b})$, where 
$\bar{q}=q_{sea}$ for each flavour, $u=u_v+u_{sea}, d=d_v+d_{sea}$ and 
$q=q_{sea}$ for all other flavours. 
The flavour structure of the light quark sea 
allows for the violation of the Gottfried sum rule. However, there is no 
information on the shape of the $\bar{d}-\bar{u}$ distribution in a fit 
to HERA data alone and so this distribution has its shape fixed consistent 
with the Drell-Yan data and its normalisation consistent 
with the size of the Gottfried sum-rule violation. 
A suppression of the strange sea with respect to the non-strange sea 
of a factor of 2 at $Q^2_0$, is also imposed
consistent with neutrino induced dimuon data from CCFR. 
Parameters are further restricted as follows.
The normalisation parameters, $p_1$, for the $d$ and $u$ valence and for the 
gluon are constrained to impose the number sum-rules and momentum sum-rule. 
The $p_2$ parameter which constrains the low-$x$ behaviour of the $u$ and $d$ 
valence distributions is set equal, 
since there is no information to constrain any difference. 
When fitting to HERA data alone it is also necessary to constrain 
the high-$x$ sea and gluon shapes, because HERA-I data do not have high 
statistics at large-$x$, in the region where these distributions are small.
The sea shape has been restricted by setting $p_4=0$ for the sea, 
but the gluon shape is constrained by including data on jet production in the 
PDF fit. Finally the ZEUS analysis has 11 free PDF parameters. 
ZEUS have included reasonable variations of 
these assumptions about the input parametrization 
in their analysis of model uncertainties. 
The strong coupling constant was fixed to $\asmz =  0.118$~\cite{lepalf}.
Full account has been taken of correlated experimental 
systematic errors by the Offset Method, 
as described in ref~\cite{zeus-s,durham}.

For the H1 analysis, the value of $Q^2_0 = 4$GeV$^2$, and 
the choice of quark distributions which are 
parametrized is different. The quarks are considered as $u$-type and $d$-type
with different parametrizations for, $xU= x(u_v+u_{sea} + c)$, 
$xD= x(d_v +d_{sea} + s)$, $x\bar{U}=x(\bar{u}+\bar{c})$ and 
$x\bar{D}=x(\bar{d}+\bar{s})$, with $q_{sea}=\bar{q}$, as 
usual, and the the form of the quark and gluon parametrizations
given by Eq.~\ref{eqn:pdf}. For $x\bar{D}$ and $x\bar{U}$ the polynomial, 
$P(x)=1.0$,
for the gluon and $xD$, $P(x)= (1+p_4 x)$, and for $xU$, 
$P(x)= (1 +p_4 x +p_5 x^3)$. The parametrization is then further restricted 
as follows.
Since the valence distributions must vanish as $x \to 0$, 
the low-$x$ parameters, $p_1$
 and $p_2$ are set equal for $xU$ and $x\bar{U}$, and for $xD$ and 
$x\bar{D}$. Since there is no information on the flavour structure of the sea 
it is 
also necessary to set $p_2$ equal for $x\bar{U}$ and $x\bar{D}$. 
The normalisation, $p_1$, of the gluon is determined from the momentum 
sum-rule and the $p_4$ parameters for $xU$ and $xD$ are determined from the 
valence number sum-rules.
Assuming that the strange and charm quark distributions can be expressed as 
$x$ independent fractions, $f_s$ and $f_c$, of the $d$ and $u$ type sea, 
gives the further constraint $p_1(\bar{U})=p_1(\bar{D}) (1-f_s)/(1-f_c)$. 
Finally there are 10 free parameters. H1 have also included reasonable 
variations of 
these assumptions in their analysis of model uncertainties. 
The strong coupling constant was fixed to $\asmz =  0.1185$ and this is 
sufficiently similar to the ZEUS choice that we can rule it out as a cause of
any significant difference. 
Full account has been taken of correlated experimental 
systematic errors by the Hessian Method, see ref.~\cite{durham}. 
 
For the ZEUS analysis, the heavy quark production scheme used is the
general mass variable flavour number scheme of Roberts and Thorne~\cite{hq}.
For the H1 analysis, the zero mass variable flavour number scheme is used. 
It is well known that these choices have a small effect on the steepness of 
the gluon at very small-$x$, such that the zero-mass choice produces a 
slightly less steep gluon. However, there is no effect on the more striking 
differences in the gluon shapes at larger $x$.

There are two differences in 
the analyses which are worth further investigation. 
The different choices for the form of the PDF parametrization at 
$Q^2_0$ and the 
different treatment of the correlated experimental uncertainties.

\begin{figure}[tbp]
\vspace{-2.0cm} 
\centerline{
\epsfig{figure=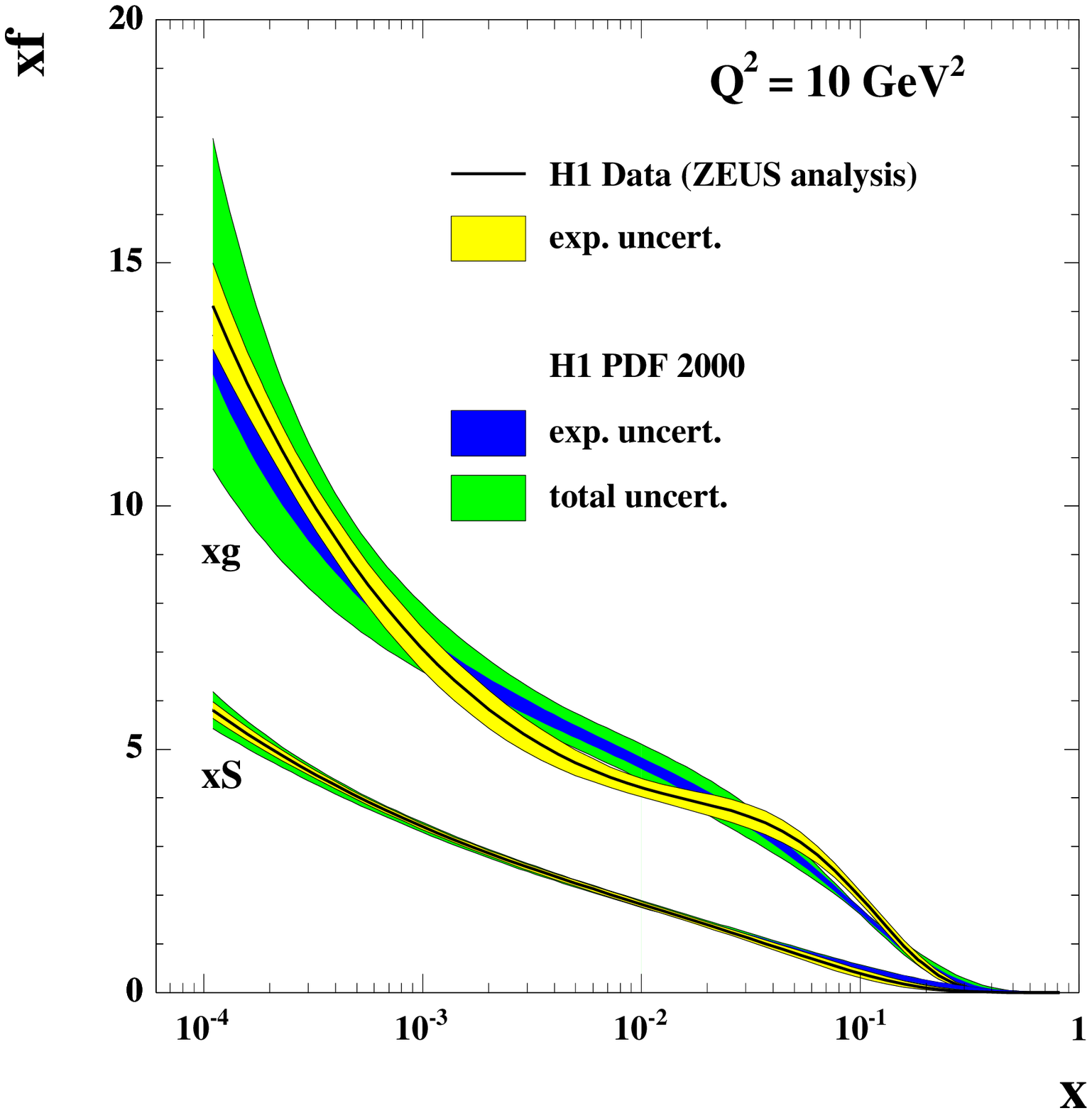,width=0.33\textwidth}
\epsfig{figure=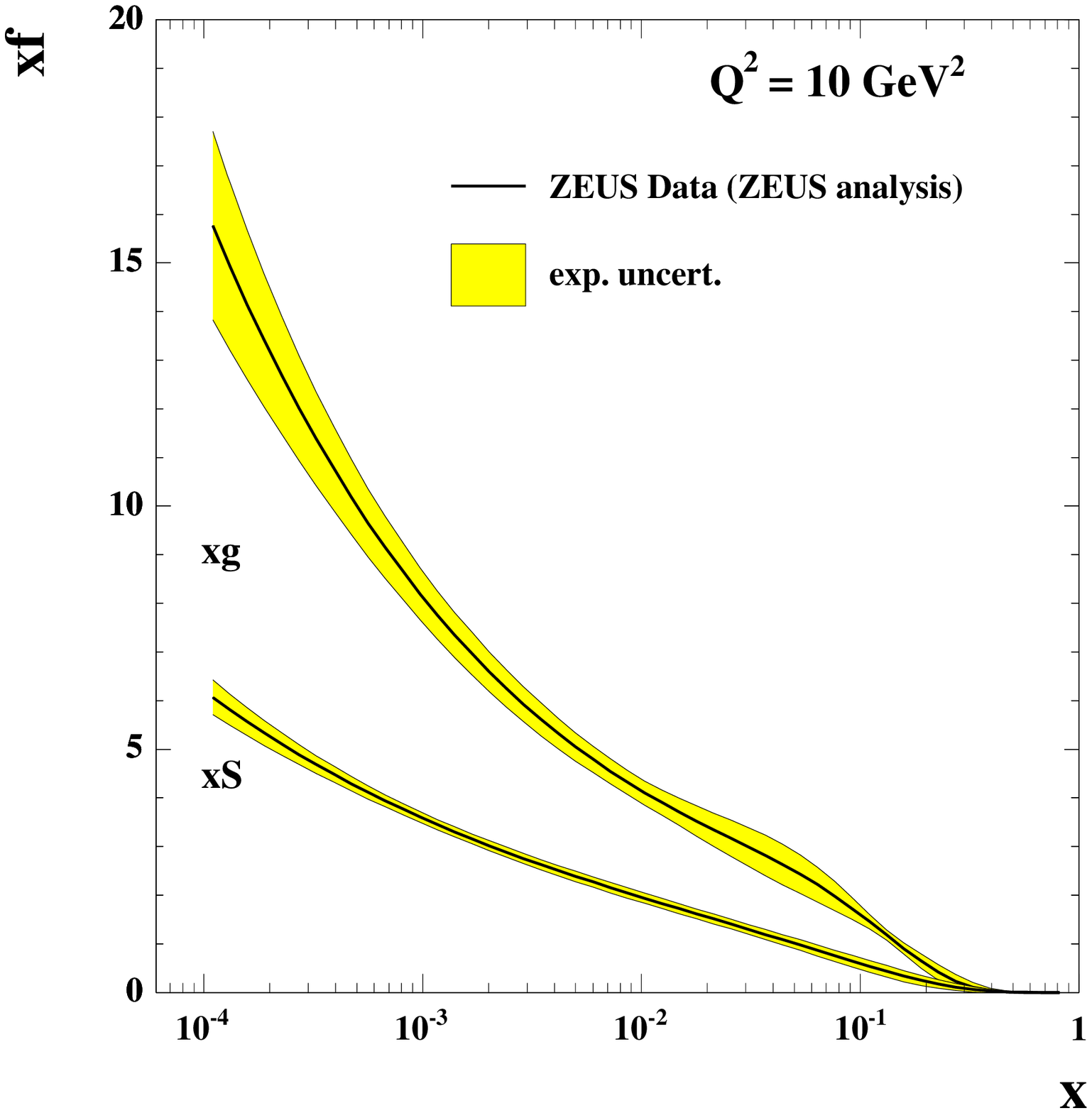,width=0.33\textwidth}
\epsfig{figure=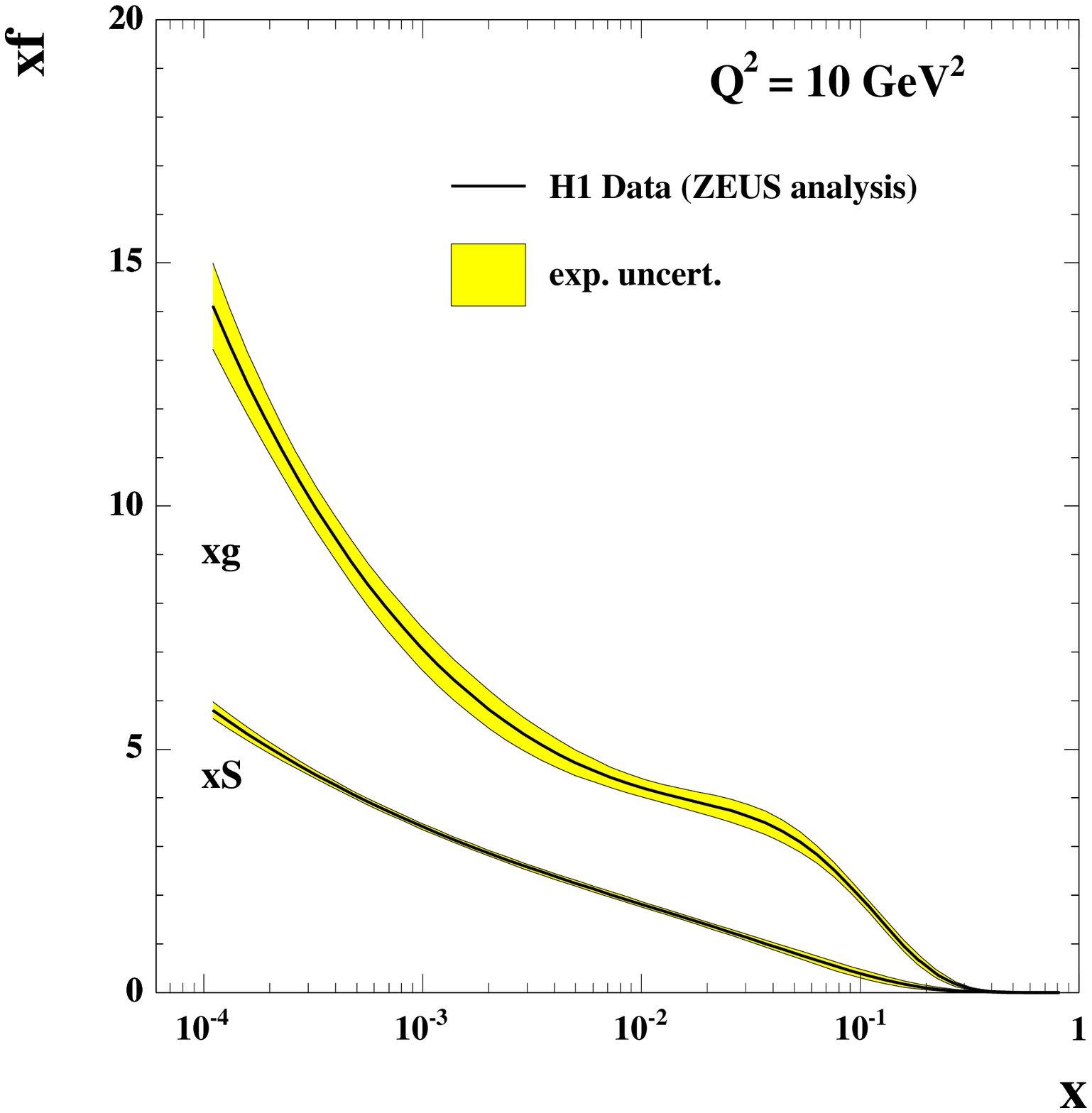,width=0.33\textwidth}
}
\caption {Sea and gluon distributions at $Q^2=10$GeV$^2$ extracted from 
different data sets and different analyses. 
Left plot: H1 data put through both ZEUS and H1 analyses.
Middle plot: ZEUS data put through ZEUS analysis. Right plot: H1 data put 
through ZEUS analysis.
}
\label{zazdzahd}
\end{figure}

So far we have compared the results of putting two different data sets into 
two different analyses. Because there are many differences in the assumptions 
going into these analyses
it is instructive to consider:(i) putting both data sets through the same analysis and (ii) putting one of the data sets through both analyses.
For these comparisons, the ZEUS analysis does NOT include the jet data, 
so that the data sets are more directly comparable, involving just the 
inclusive double differential cross-section data. 
Fig.~\ref{zazdzahd} compares the sea and gluon PDFs, 
at $Q^2=10$GeV$^2$, extracted from H1 data using the H1 PDF analysis 
with those extracted from H1 data using the ZEUS PDF analysis. 
These alternative analyses of the same data 
set give results which are compatible within the model dependence error 
bands. Fig.~\ref{zazdzahd} also compares the sea and gluon PDFs extracted from 
ZEUS data using the ZEUS analysis with those extracted from H1 data using 
the ZEUS analysis. From 
this comparison we can see that the different data sets lead to somewhat 
different gluon shapes even when put through exactly the same analysis. 
Hence the most of the difference in shape of the ZEUS and H1 PDF analyses can 
be traced back to a difference at the level of the data sets.


Before going further it is useful to discuss the treatment of correlated 
systematic errors in the ZEUS and H1 analyses. A full discussion of the 
treatment of correlated systematic errors in PDF 
analyses is given in ref~\cite{dcs}, only salient details are recapped here. 
Traditionally, experimental collaborations have evaluated an overall systematic
uncertainty on each data point and these have been treated as uncorrelated, 
such that they are simply added to the statistical uncertainties in quadrature
when evaluating $\chi^2$. However, modern deep inelastic scattering experiments
have very small statistical uncertainties, so that the contribution of 
systematic uncertainties becomes dominant and consideration of 
point to point correlations between systematic uncertainties is essential.

For both ZEUS and H1 analyses
the formulation of the $\chi^2$ including correlated systematic uncertainties
 is constructed as follows. The correlated uncertainties
are included in the theoretical prediction, $F_i(p,s)$, such that
\[ 
F_i(p,s) = F_i^{\rm NLOQCD}(p) + 
\sum_{\lambda} s_{\lambda} \Delta^{\rm sys}_{i\lambda}
\]
where, $F_i^{\rm NLOQCD}(p)$, represents the prediction 
from NLO QCD in terms of the theoretical parameters $p$,
and the parameters $s_\lambda$ represent independent variables 
for each source of
 systematic uncertainty. They have zero mean and unit variance by construction.
The symbol 
$\Delta^{\rm sys}_{i\lambda}$ represents the one standard deviation correlated 
systematic error on data point $i$ due to correlated error 
source $\lambda$.
The $\chi^2$ is then formulated as 
\begin{equation}
\chi^2 = \sum_i \frac{\left[ F_i(p,s)-F_i(\rm meas) \right]^2}{\sigma_i^2} + \sum_\lambda s^2_\lambda 
\label{eq:chi2}
\end{equation}
where, $F_i(\rm meas)$, represents a measured data point and the symbol 
$\sigma_i$ represents the one standard deviation uncorrelated 
error on data point $i$, from both statistical and systematic sources. 
The experiments use this $\chi^2$ in different ways. ZEUS uses the Offset 
method and H1 uses the Hessian method.
 
Traditionally, experimentalists have used `Offset' methods to account for
correlated systematic errors. The $\chi^2$ is formluated without any terms
due to correlated systematic errors ($s_\lambda=0$ in Eq.~\ref{eq:chi2}) for
evaluation of the central values of the fit parameters. 
However, the data points are then offset to account for each 
source of systematic error in turn 
(i.e. set $s_\lambda = + 1$ and then $s_\lambda = -1$ for each source 
$\lambda$) 
and a new fit is performed for each of these
variations. The resulting deviations of the theoretical parameters 
from their  central  values are added in 
quadrature. (Positive and  negative deviations are added 
in quadrature separately.) This method does not assume that the systematic
uncertainties are Gaussian distributed.
An equivalent (and much more efficient) procedure to perform the
Offset method has been given by 
Pascaud and Zomer~\cite{pz}, and this is what is actually used.
The Offset method is a conservative method of error estimation  
as compared to the Hessian method. 
It gives fitted theoretical predictions which are as close as 
possible to the central values of the published data. It does not use the full 
statistical power of the fit to improve the estimates of $s_\lambda$, 
since it choses to mistrust the systematic error estimates,
but it is correspondingly more robust.

The Hessian method is an alternative procedure in which the systematic
uncertainty parameters $s_\lambda$ are allowed to vary in the main fit 
when determining the values of the theoretical parameters. 
Effectively, the theoretical prediction is not fitted
to the central values of the published experimental data, but  
these data points are allowed to move
collectively, according to their correlated systematic uncertainties.
 The theoretical prediction determines the 
optimal settings for correlated systematic shifts of experimental data points 
such that the most consistent fit to all data sets is obtained. Thus, 
in a global fit, systematic shifts in 
one experiment are correlated to those in another experiment by the fit.
In essence one is allowing the theory to calibrate the detectors. This requires
great confidence in the theory, but more significantly, it requires confidence
in the many model choices which go into setting the boundary conditions for 
the theory (such as the parametrization at $Q^2_0$). 

\begin{figure}[tbp]
\vspace{-2.0cm} 
\centerline{
\epsfig{figure=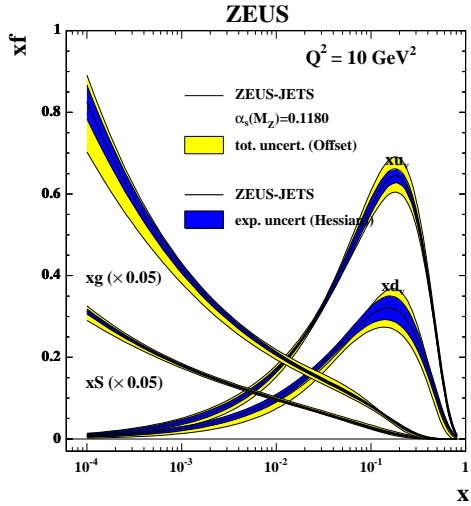,height=7cm}
}
\caption {PDFs at $Q^2=10$GeV$^2$, for the ZEUS analysis of ZEUS data 
performed by the Offset and the Hessian methods.
}
\label{fig:offhess}
\end{figure}

The ZEUS analysis can be performed using the Hessian method as well as the 
Offset method and
Fig.~\ref{fig:offhess} compares the PDFs, and their uncertainties, 
extracted from ZEUS data using these two methods. 
The central values of the different methods are in good agreement but 
the use of the Hessian method results in smaller uncertainties, for a 
the standard set of model assumptions, since the input data can be shifted 
within their correlated systematic uncertainties to suit the theory better. 
However, model uncertainties are more significant for the Hessian method than 
for the Offset method. The experimental 
uncertainty band for any one set of model choices 
is set by the usual $\chi^2$ tolerance, $\Delta \chi^2=1$, but the 
acceptability of a different set of choices is judged by the hypothesis 
testing criterion, such that
the $\chi^2$ should be approximately in the range $N \pm \surd (2N)$,
where $N$ is the number of degrees of freedom. The PDF
parameters obtained for the different model choices can differ 
by much more than their experimental uncertainties, because each model choice
can result in somewhat 
different values of the systematic uncertainty parameters, $s_\lambda$, and 
thus a different estimate of the shifted positions of the data points. This 
results in a larger spread of model uncertainty than in the Offset method, 
for which the data points cannot move. Fig~\ref{fig:h1zeus} illustrates the 
comparability of the ZEUS (Offset) total uncertainty  
estimate to the H1 (Hessian) experimental 
plus model uncertainty estimate. 

Another issue which arises in relation to the Hessian method is that
the data points should not be shifted far outside their 
one standard deviation systematic uncertainties. This can indicate 
inconsistencies between data sets, or parts of data sets, with respect to 
the rest of the data. The CTEQ collaboration have considered data
inconsistencies in their most recent global fit~\cite{cteq}.
They use the Hessian method but they increase the resulting uncertainty 
estimates, by 
increasing the $\chi^2$ tolerance to $\Delta \chi^2 = 100$, to allow for both 
model uncertainties and data inconsistencies. 
 In setting this tolerance they have considered the distances
from the $\chi^2$-minima of individual data sets 
to the global minimum for all data sets. These distances
 by far exceed the range allowed by the $\Delta \chi^2 =1$ criterion.
Strictly speaking such variations can indicate that data sets are inconsistent
but the CTEQ collaboration take the view 
that all of the current 
world data sets must be considered acceptable and compatible at some level,
even if strict statistical criteria are not met, since the conditions
for the application of strict criteria, namely Gaussian error distributions, 
are also not met. It is not possible to simply drop ``inconsistent'' data 
sets, as then the partons in some regions would lose important constraints.
On the other hand the level of ``inconsistency'' should 
be reflected in the uncertainties of the PDFs.
This is achieved by raising the $\chi^2$ tolerance. This 
results in uncertainty estimates which are comparable to 
those achieved by using the Offset method~\cite{durham}.

\begin{figure}[tbp]
\centerline{
\epsfig{figure=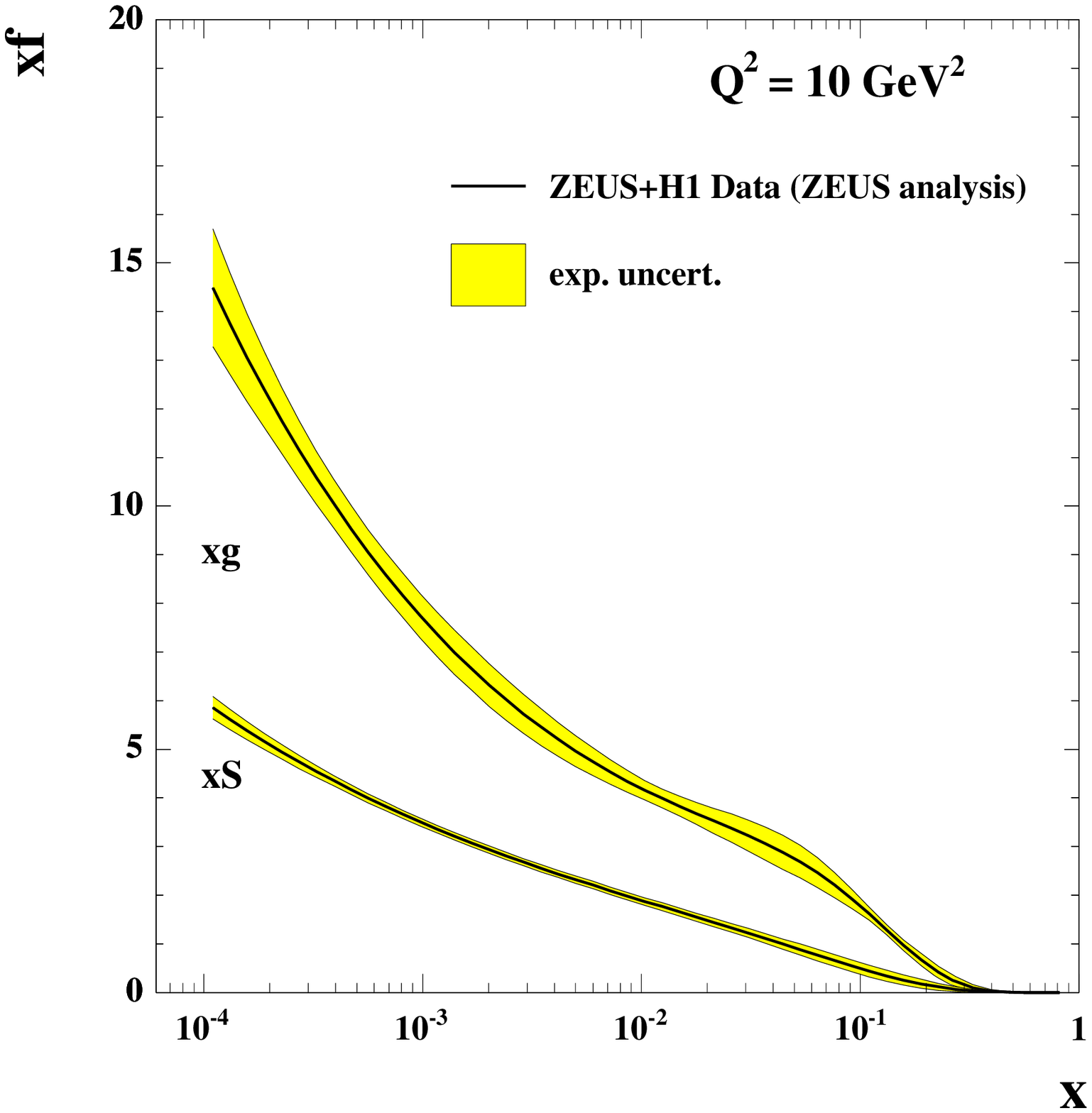,height=5cm}
\epsfig{figure=SEAGLU_ZEUSData_ZEUSANAL.eps,height=5cm}}
\centerline{
\epsfig{figure=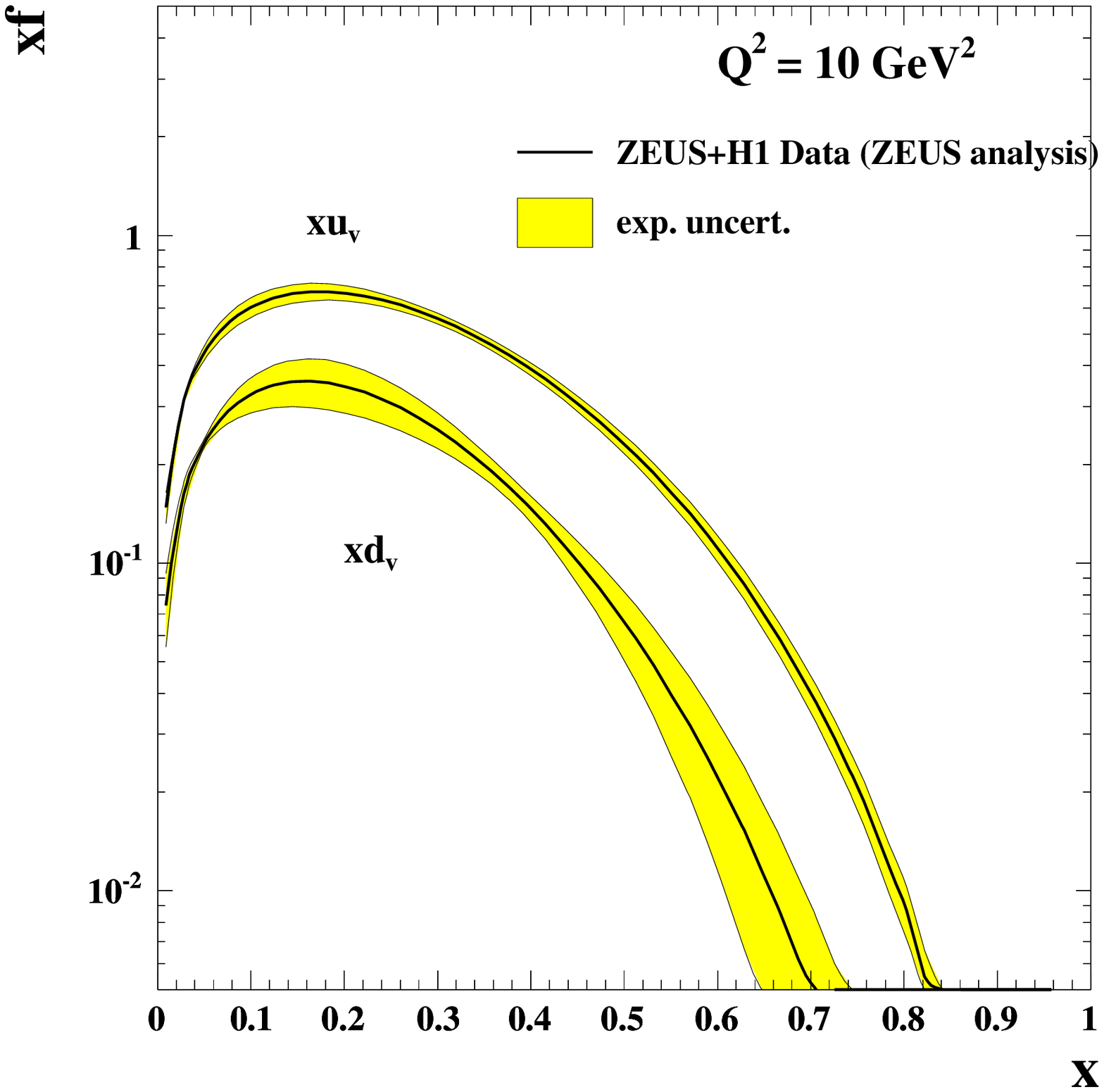,height=5cm}
\epsfig{figure=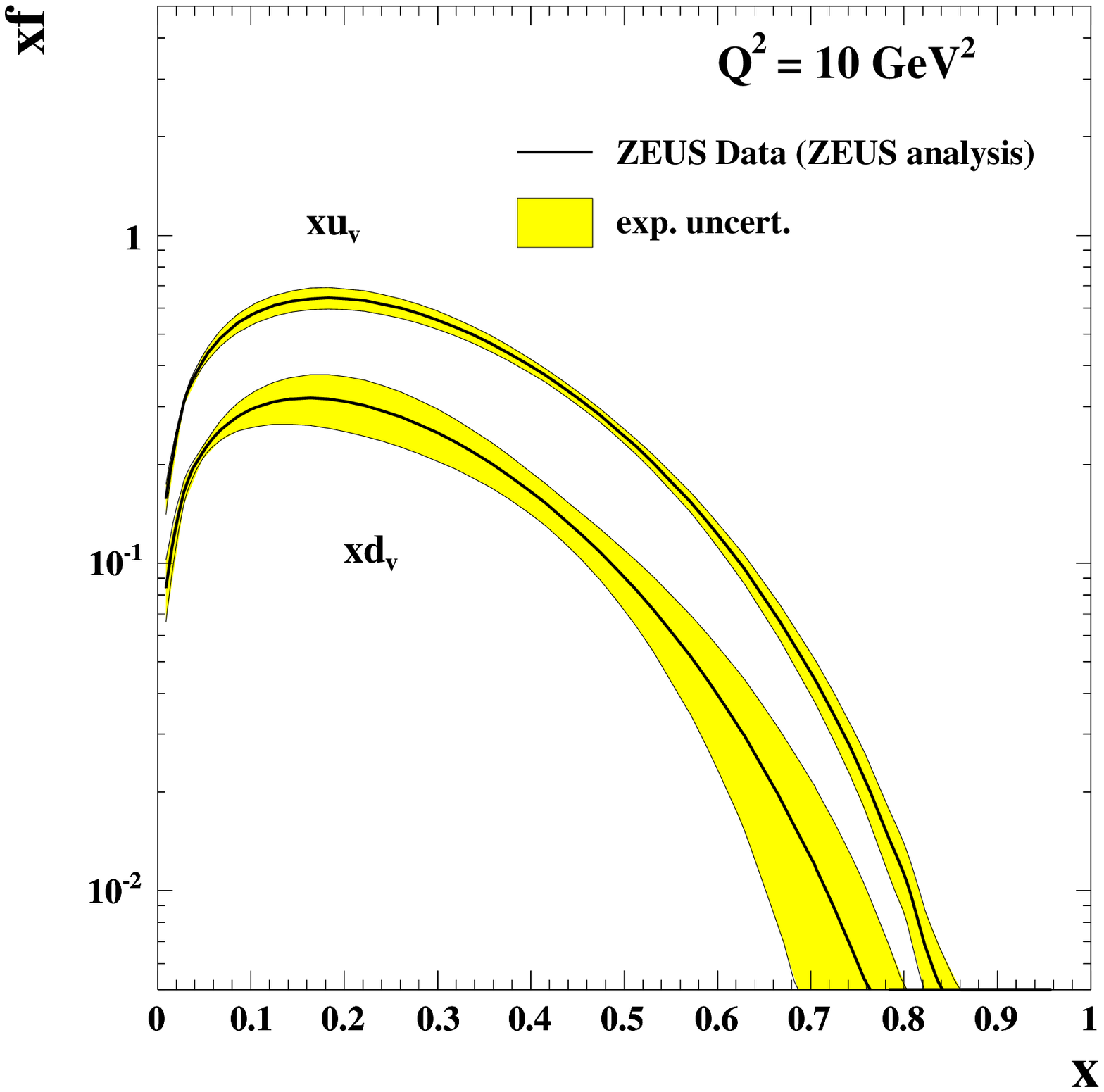,height=5cm}
}
\caption {Top plots: Sea and gluon distributions at $Q^2=10$GeV$^2$ extracted 
from H1 and ZEUS data using the ZEUS analysis (left) compared to those 
extracted from ZEUS data alone using the ZEUS analysis (right).
Bottom Plots: Valence distributions at $Q^2=10$GeV$^2$, extracted from 
H1 and ZEUS data using the ZEUS analysis (left) compared to those extracted 
from ZEUS data alone using the ZEUS analysis (right). 
}
\label{fig:zh1tog}
\end{figure}

Using data from a single experiment avoids questions of data consistency, 
but to
get the most information from HERA
it is necessary to put ZEUS and H1 data sets into the same 
analysis together, and then questions of consistency arise. 
Fig~\ref{fig:zh1tog} compares the sea and gluon PDFs and the 
$u$ and $d$ valence PDFs extracted from the ZEUS PDF 
analysis of ZEUS data alone,
to those extracted from the ZEUS PDF analysis of both H1 and ZEUS data.
It is noticeable that, for the low-$x$ sea and gluon PDFs, combining the 
data sets does not bring a reduction in uncertainty
equivalent to doubling the statistics. This is because the data which 
determine these PDFs are systematics limited. In fact there is some degree of 
tension between the ZEUS and the H1 data sets, such that the $\chi^2$ per 
degree of freedom rises for both data sets when they are fitted together. 
The Offset method of treating the systematic errors 
reflects this tension such that the overall uncertainty is not much improved 
when H1 data are added to ZEUS data. However, the uncertainty on the high-$x$
valence distributions is reduced by the input of 
H1 data, since the data are still statistics limited at high $x$.

\begin{figure}[tbp]
\centerline{
\epsfig{figure=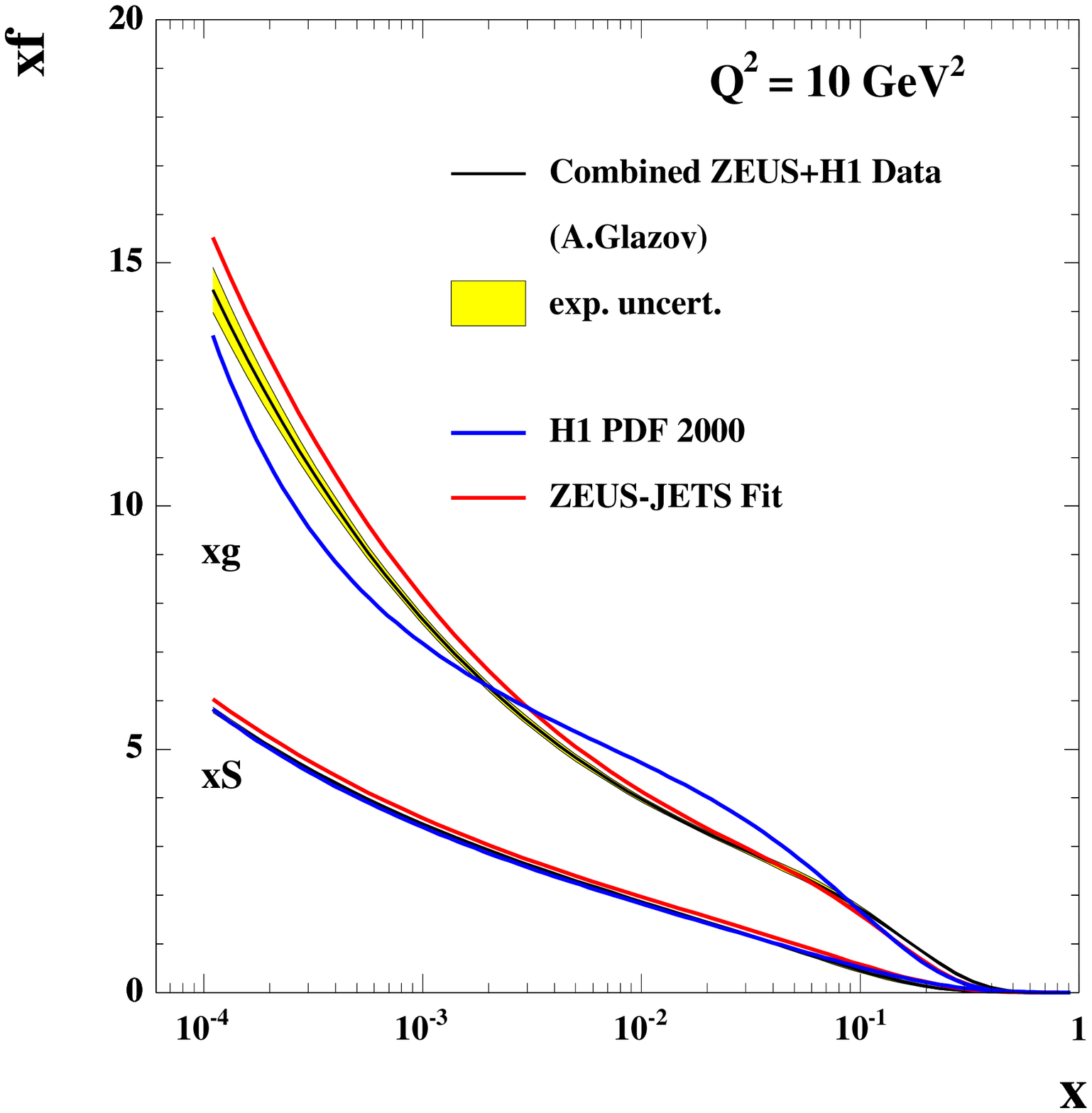,height=5cm}
\epsfig{figure=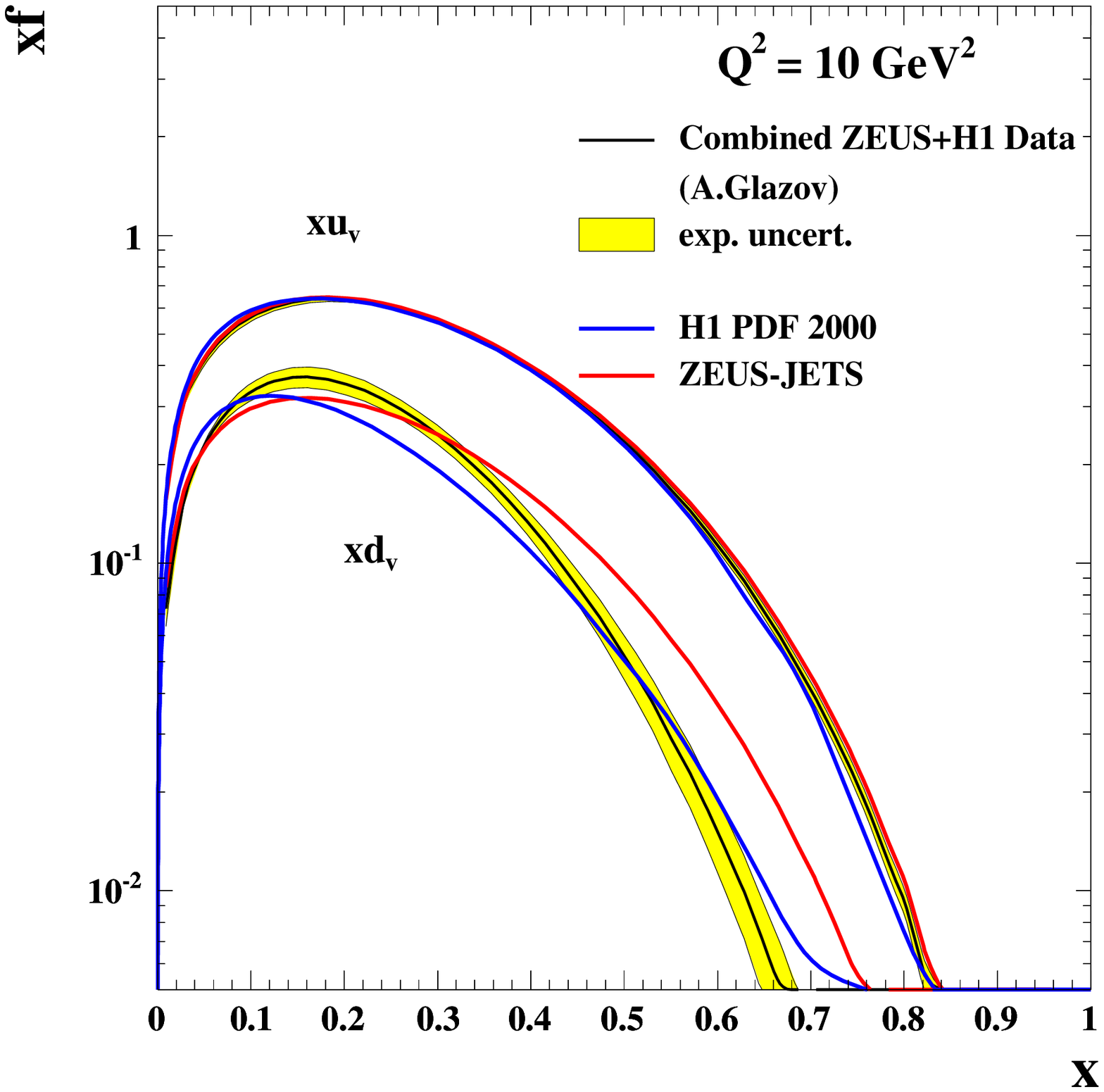,height=5cm}}

\caption {Left plot: Sea and gluon distributions at $Q^2=10$GeV$^2$, extracted 
from the combined H1 and ZEUS data set using the ZEUS analysis.
Right plot: Valence distributions at $Q^2=10$GeV$^2$, extracted from 
the combined H1 and ZEUS data set using the ZEUS analysis.
}
\label{fig:glazov}
\end{figure}

Thus there could be an advantage in combining ZEUS and H1  
data in a PDF fit if the tension between the data 
sets could be resolved. It is in this context the question of combining these 
data into a single data set arises. The procedure for combination is detailed 
in the contribution of S. Glazov to these proceedings.
Essentially, since ZEUS and H1 are measuring the same physics in the same 
kinematic region, one can try to combine them using a 'theory-free' 
Hessian fit in which the only assumption is that there is a true 
value of the cross-section, for each process, at each $x,Q^2$ point. 
The systematic uncertainty parameters, $s_\lambda$, of each experiment 
are fitted to determine the best fit to this assumption. 
Thus each experiment is calibrated to the other. This works well because the 
sources of systematic uncertainty in each experiment are rather different. 
Once the procedure has been performed the resulting systematic uncertainties 
on each of the combined data points are significantly smaller than the 
statistical errors. Thus one can legitimately make a fit to the combined data 
set in which these statistical and systematic uncertainties are simply 
combined in quadrature. The result of making such a fit, using the ZEUS 
analysis, is shown in Fig.~\ref{fig:glazov}.
The central values of the ZEUS and H1 published 
analyses are also shown for comparison.
Looking back to Fig.~\ref{fig:zh1tog} one can see that there has been a 
dramatic reduction in the level of uncertainty 
compared to the ZEUS Offset method fit to the separate ZEUS and H1 data sets.
This result is very promising. A preliminary study of model dependence, varying
the form of the polynomial, $P(x)$, used in the PDF paremtrizations at $Q^2_0$,
also indicates that model dependence is relatively small. 

\begin{figure}[tbp]
\centerline{
\epsfig{figure=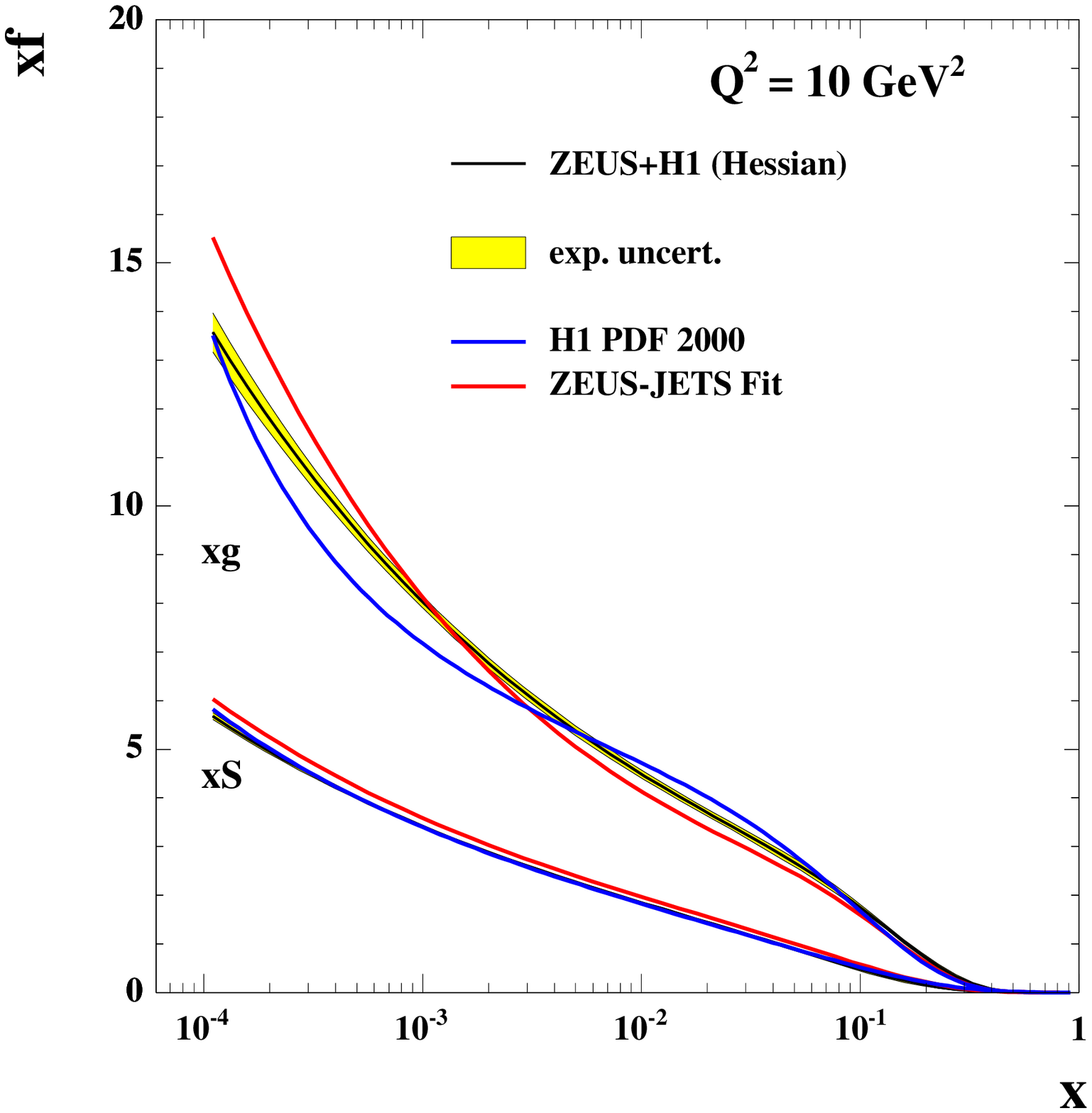,height=5cm}
\epsfig{figure=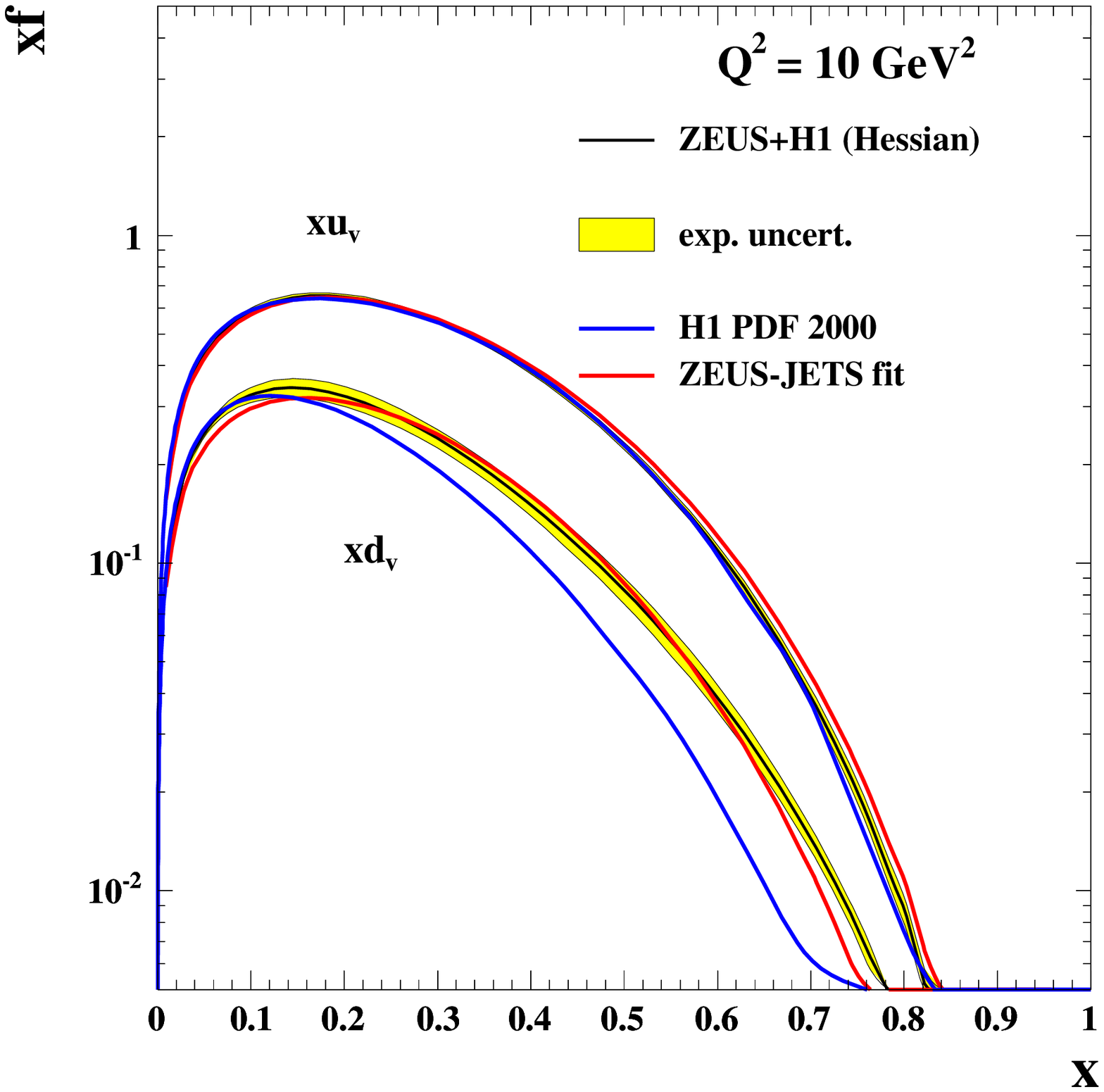,height=5cm}}

\caption {Left plot: Sea and gluon distributions at $Q^2=10$GeV$^2$, extracted
from the H1 and ZEUS data sets using the ZEUS analysis done by Hessian
method.
Right plot: Valence distributions at $Q^2=10$GeV$^2$, extracted
from the H1 and ZEUS data sets using the ZEUS analysis done by Hessian
method. 
}
\label{fig:zh1hess}
\end{figure}

The tension between ZEUS and H1 data could have been resolved by
putting them both into a PDF fit using the Hessian method to shift the data 
points. That is, rather than calibrating the two experiments to each other in 
the 'theory-free' fit, we could have used the theory of 
pQCD to calibrate each experiment. Fig.~\ref{fig:zh1hess} shows the PDFs 
extracted when the ZEUS and H1 data sets are put through the ZEUS PDF analysis
procedure using the Hessian method. 
The uncertainties on the resulting PDFs are comparable to those found for the 
fit to the combined data set, see Fig.~\ref{fig:glazov}. 
However, the central values of the resulting 
PDFs are rather different- particularly for the less well known gluon and
$d$ valence PDFs. For both of the fits shown in 
Figs.~\ref{fig:glazov},~\ref{fig:zh1hess} the values of the systematic error
 parameters, $s_\lambda$, for each experiment have been allowed to float so 
that the data points are shifted to give a better fit to our assumptions, but 
the values of the systematic error parameters chosen 
by the 'theory-free' fit and by the PDF fit are rather different. A 
representaive sample of these values is given in 
Table~\ref{tab:sl}. These discrepancies might be somewhat alleviated by a full 
consideration of model errors in the PDF fit, or of appropriate $\chi^2$ 
tolerance when combining the ZEUS and H1
experiments in a PDF fit, but these differences should make us wary about the 
uncritical use of the Hessian method.
\begin{table}[t]
\begin{tabular}{ccc}\\
 \hline
Syatematic uncertainty $s_\lambda$  &  in PDF fit&  in Theory-free fit \\
 \hline
 ZEUS electron efficiency  & 1.68 & 0.31 \\
 ZEUS electron angle  & -1.26 & -0.11 \\
 ZEUS electron energy scale  & -1.04 & 0.97 \\
 ZEUS hadron calorimeter energy scale  & 1.05 & -0.58 \\
 H1 electron energy scale  & -0.51 & 0.61 \\
 H1 hadron energy scale  & -0.26 & -0.98 \\
 H1 calorimeter noise  & 1.00 & -0.63 \\
 H1 photoproduction background  & -0.36 & 0.97 \\

 \hline\\
\end{tabular}
\caption{Systematic shifts for ZEUS and H1 data as determine by a joint pQCD 
PDF fit, and as determined by the theory-free data combination fit}
\label{tab:sl}
\end{table}

\bibliographystyle{heralhc} 
{\raggedright
\bibliography{heralhc}

\providecommand{\etal}{et al.\xspace}
\providecommand{\coll}{Coll.}
\catcode`\@=11
\def\@bibitem#1{%
\ifmc@bstsupport
  \mc@iftail{#1}%
    {;\newline\ignorespaces}%
    {\ifmc@first\else.\fi\orig@bibitem{#1}}
  \mc@firstfalse
\else
  \mc@iftail{#1}%
    {\ignorespaces}%
    {\orig@bibitem{#1}}%
\fi}%
\catcode`\@=12
\begin{mcbibliography}{10}

\bibitem{mrst}
A.D.~Martin et al.,
\newblock Eur. Phys.J{} {\bf C23},~73~(2002)\relax
\relax
\bibitem{cteq}
J.~Pumplin et al.,
\newblock JHEP{} {\bf 0207},~012~(2002)\relax
\relax
\bibitem{zeus-s}
ZEUS Coll., S.~Chekanov et al.,
\newblock Phys. Rev{} {\bf D~67},~012007~(2003)\relax
\relax
\bibitem{zeusj}
ZEUS Coll., S.~Chekanov et al.,
\newblock Eur.Phys.J{} {\bf C~42},~1~(2005)\relax
\relax
\bibitem{h1}
H1 Coll., C.Adloff et al.,
\newblock Eur.Phys.J{} {\bf C~30},~32~(2003)\relax
\relax
\bibitem{ap}
G. Altarelli, G.~Parisi,
\newblock Nucl.Phys.{} {\bf B126},~298~(1977)\relax
\relax
\bibitem{gl}
V.N.~Gribov, L.N.~Lipatov,
\newblock Sov.J.Nucl.Phys{} {\bf 15},~438~(1972)\relax
\relax
\bibitem{l}
L.N.~Lipatov,
\newblock Sov.J.Nucl.Phys{} {\bf 20},~94~(1975)\relax
\relax
\bibitem{d}
Yu.L.~Dokshitzer,
\newblock JETP{} {\bf 46},~641~(1977)\relax
\relax
\bibitem{dcs}
R C E Devenish and A M Cooper-Sarkar,
\newblock {\em Deep Inelastic Scattering}.
\newblock Oxford Unviersity Press, Oxford, 2004\relax
\relax
\bibitem{lepalf}
S.~Eidelman,
\newblock Phys.Lett{} {\bf B~592},~1~(2004)\relax
\relax
\bibitem{durham}
A.M.~Cooper-Sarkar,
\newblock J.Phys{} {\bf G~28},~2669~(2002)\relax
\relax
\bibitem{hq}
R.S.~Thorne and R.G.~Roberts,
\newblock Phys.Rev{} {\bf D57},~6871~(1998)\relax
\relax
\bibitem{pz}
C. Pascaud and F. Zomer,
\newblock {\em Correlated systematic error propagation}.
\newblock Preprint \mbox{LAL-95-05}, 1995\relax
\relax
\end{mcbibliography}
}
\end{document}